\documentclass[pra,twocolumn,showpacs,floatfix]{revtex4}

\usepackage{setspace}
\usepackage{amsmath,amsthm,amsfonts,amssymb,bm,bbm}
\usepackage{epsfig}
\usepackage{times}
\usepackage{psfrag}
\usepackage[dvips]{color}

\newcommand{\mc}{ {m^{}_C} }
\newcommand{\me}{ {m^{}_E} }
\newcommand{\mr}{ {m^{}_R} }

\newcommand{\dind}{d_{\rm ind}}
\newcommand{\udes}{ L_S }

\newcommand{\ns}{{n^{}_S}}
\newcommand{\nc}{{n^{}_C}}

\newcommand{\nca}{{n^{}_{CA}}}
\newcommand{\nra}{{n^{}_{RA}}}

\newcommand{\normf}[1]{\|#1\|_{\rm Fro}}

\newcommand{\Is}{I_S}
\newcommand{\Ic}{I_C}

\newcommand{\Ie}{I_E}

\newcommand{\favg}{f_{\rm avg}}

\newcommand{\adj}{{\dagger}}

\newcommand{\btab}{\begin{tabular}}
\newcommand{\etab}{\end{tabular}}

\newcommand{\rhoc}{\rho^{}_{\rm C}}
\newcommand{\sigc}{\sig^{}_{\rm C}}
\newcommand{\rhos}{\rho^{}_{\rm S}}
\newcommand{\rhohs}{\hat{\rho}^{}_{\rm S}}

\newcommand{\ket}[1]{\mbf{|}#1\mbf{\rangle}}
\newcommand{\bra}[1]{\mbf{\langle}#1\mbf{|}}

\newcommand{\Rcal}{{\cal R}}

\newcommand{\trace}{{\bf Tr}}

\newcommand{\real}{{\rm Re}}

\newcommand{\alf}{\alpha}

\newcommand{\lam}{\lambda}

\renewcommand{\th}{\theta}

\newcommand{\sig}{\sigma}

\newcommand{\Del}{\Delta}
\newcommand{\Gam}{\Gamma}

\newcommand{\norm}[1]{ \left\| #1 \right\| }

\newcommand{\rhoh}{\hat{\rho}}

\newcommand{\Cb}{\bar{C}}

\newcommand{\Ccal}{{\mathcal C}}

\newcommand{\Ecal}{{\mathcal E}}

\newcommand{\eg}{\emph{e.g.}}
\newcommand{\ie}{\emph{i.e.}}

\newcommand{\bquem}{\begin{quote}\begin{em}}
\newcommand{\equem}{\end{em}\end{quote}}

\newcommand{\blist}{\begin{description}}
\newcommand{\elist}{\end{description}}

\newcommand{\bquote}{\begin{quote}}
\newcommand{\equote}{\end{quote}}

\newcommand{\ben}{\begin{enumerate}}
\newcommand{\een}{\end{enumerate}}

\newcommand{\bit}{\begin{itemize}}
\newcommand{\eit}{\end{itemize}}

\newcommand{\bea}{\begin{array}}
\newcommand{\eea}{\end{array}}

\newcommand{\bds}{\begin{displaystyle}}
\newcommand{\eds}{\end{displaystyle}}

\newcommand{\mbf}[1]{\mbox{\boldmath $#1$}}

\newcommand{\refeq}[1]{(\ref{eq:#1})}

\newcommand{\set}[2]{ \left\{ \,#1\, \left| \,#2\, \right.\right\} }

\newcommand{\seq}[1]{ \left\{ #1 \right\} }

\newcommand{\opt}{{\rm opt}}

\makeatletter
\def\beq{\@ifnextchar 
[{\@tempswatrue\@beq}{\@tempswafalse\@beq[]}}
\def\@beq[#1]{\begin{equation}\edef\@tmparg{#1}\ifx\@tmparg\@e
mpty \else
	\label{#1}\fi}
\newcommand{\eeq}{\end{equation}}
\makeatother 

\newcommand{\beqaa}{\begin{eqnarray*}}
\newcommand{\eeqaa}{\end{eqnarray*}}

\newcommand{\beqa}{\begin{eqnarray}}
\newcommand{\eeqa}{\end{eqnarray}}

\newcommand{\bc}{\begin{center}}
\newcommand{\ec}{\end{center}}

\renewcommand{\normf}[1]{\|#1\|_{\rm F}}
\newcommand{\algsdp}{Algorithm-1}
\newcommand{\algcls}{Algorithm-2}
\newcommand{\av}{{\rm avg}}

\renewcommand{\favg}{f}
\renewcommand{\dind}{d}

\newcommand{\Xcal}{{\cal X}}

\begin{document}

\title{Channel-Optimized Quantum Error Correction}
\author{Soraya Taghavi}
\affiliation{Department of Electrical Engineering
  and the Center for Quantum Information Science and Technology,
  University of Southern California, Los Angeles, CA 90089}
\author{Robert L. Kosut}
\affiliation{
SC Solutions, Systems \& Control Division, 1261 Oakmead
Pkwy., Sunnyvale, CA 94085}
\author{Daniel A. Lidar}
\affiliation{
Departments of Electrical Engineering,
Chemistry, and Physics, and the Center for Quantum Information Science and
Technology, University of Southern California, Los Angeles, CA 90089}

\begin{abstract}

We develop a theory for finding quantum error correction (QEC)
procedures which are optimized for given noise channels. Our theory
accounts for uncertainties in the noise channel, against which our QEC
procedures are robust. We demonstrate via numerical examples that our
optimized QEC procedures always achieve a higher channel fidelity
than the standard error correction method, which is agnostic about the
specifics of the channel. This demonstrates the importance of channel
characterization before QEC procedures are applied. Our main novel
finding is that in the setting of a known noise channel the recovery
ancillas are redundant for optimized quantum error correction. We show
this using a general rank minimization heuristic and supporting
numerical calculations. Therefore, one can further improve the
fidelity by utilizing all the available ancillas in the encoding
block.
\end{abstract}

\maketitle

\section{Introduction}

\label{sec:intro}

Quantum error correction is often considered the
backbone of quantum information processing, since it converts what is
essentially an analog information processor, subject to a continuum of
errors, into a digital one, whose errors are discretized. The theory
of quantum error correction was developed in analogy to classical
coding for noisy channels
\cite{Shor:95,Gott:96,Steane:96,Laflamme:96,KnillL:97,NielsenC:00}. These
initial efforts focused on finding conditions and procedures for
perfect recovery of quantum states passing through noisy channels.
Recently, several authors considered error correction design as an
optimization problem, with fidelity as the optimization target
\cite{ReimpellW:05,YamamotoHT:05,KosutL:06,KosutAL:06}. In this work
we further develop the theory of optimal quantum error correction.
As in \cite{ReimpellW:05,YamamotoHT:05,KosutL:06,KosutAL:06},
we consider the scenario where one has knowledge of the noise channel,
and find correspondingly optimal codes. That is, we assume that one
has already performed a channel identification procedure, e.g., via
quantum process tomography \cite{tom}. We show how, armed with a
knowledge of the channel, one can design highly robust error
correction procedures, whose fidelity is always at least as good as
that of the ``agnostic'' codes of standard error correction
\cite{Shor:95,Gott:96,Steane:96,Laflamme:96,KnillL:97,NielsenC:00}.

More specifically, we present an indirect approach to fidelity
maximization based on minimizing the error between the actual channel
and the desired channel. This approach, like the previously developed approach to direct fidelity
optimization,
leads naturally to bi-convex optimization problems, namely, two
semidefinite programs (SDPs) \cite{BoydV:04} which can be iterated
between the recovery and encoding. For a given encoding the problem is
convex in the recovery. For a given recovery, the problem is convex in
the encoding. An important advantage of this approach is that noisy
channels, which do not satisfy the standard assumptions for perfect
correction
\cite{Shor:95,Gott:96,Steane:96,Laflamme:96,KnillL:97,NielsenC:00},
can be optimized for the best possible encoding and recovery.

The conventional fidelity optimization targets are the encoding and
recovery operators. An important way in which the present work differs
from previous studies is in the fact that we further add the
distribution of the ancillas in the encoding and recovery to the
optimization problem. This way, we utilize all possible degrees of
freedom for optimization. As a consequence, we find a rather
surprising result: in the optimized error correction procedure the
fidelity is indifferent to the existence of the recovery
ancillas. This result paves the way toward a more efficient
utilization of the ancillas. Namely, we can use all the available
ancilla qubits in the encoding to increase the fidelity.

Standard error correction schemes, as well as those produced by the
aforementioned optimization methods which are tuned to specific
errors, are often not robust to even small changes in the error
channel. These errors can be mitigated by fault-tolerant methods which
rely on several levels of code concatenation
\cite{Gaitan:book}. However, our method naturally enjoys a desirable
robustness against error variations.  We show a means to incorporate
specific models of error channel uncertainty, resulting in highly
robust error correction.
Nevertheless, concatenated fault tolerant quantum error correction
still enjoys a certain important advantage over the procedures we
derive in this work, namely, it is robust also against imperfections
in the encoding and recovery procedures, while 
here, as is the case for standard quantum error correction formulations,
we assume these to be perfectly executed.

Since the number of optimization variables scales exponentially with
the number of qubits used in the encoding and recovery operations, the
computational effort required to solve any of the semidefinite program
optimization (SDP) problems is similarly burdened. In order to reduce
this effort we propose an approach based on optimization via the
constrained least squares method. This alternative approach for
solving the indirect optimization problem does not utilize
semidefinite programming, and is
significantly faster in our numerical simulations.  Surprisingly, this
method returns the exact same result as the SDP approach.

The organization of the paper is as follows. In Section
\ref{sec:problem}, we explain the problem formulation including the
standard error correction model and state the direct and indirect
optimization problems to be addressed. The indirect approach is
explored in Section \ref{sec:indirect}. In Section \ref{sec:ancilla}, we
investigate the optimal distribution of the ancillas between the
recovery and encoding. Examples of the methods presented are given in
Section \ref{sec:examples}. Appendices \ref{sec:maxvr01}-\ref{sec:copt}
provide proofs and supporting material.

\section{Problem formulation}

\label{sec:problem}

\subsection{Standard error correction model}

Subject to
the assumption that the initial system-bath
state is classically correlated
\cite{ShabaniLidar:08}, the dynamics of an open quantum
system can be represented in an elegant form known as the Kraus
\emph{operator sum representation} (OSR). In this representation, the
noise $\Ecal$ is described in terms of a completely-positive (CP) map:
$\rho \to \sum_i A_i \rho A_i^\dagger$ \cite{NielsenC:00}. Here $\rho$
is the initial system density matrix and the operators $A_i$, known as
Kraus operators, or operation elements, satisfy the normalization
relation $\sum_i A_i^\dagger A_i = I$ (identity). The standard error
correction procedure involves CP encoding ($\Ccal$), error ($\Ecal$),
and recovery ($\Rcal$) maps (or channels): $\rhos\overset{\Ccal
}{\to}\rhoc\overset{\Ecal}{\to }\sigc \overset{\Rcal}{\to }\rhohs$, as
shown pictorially in the block diagram of Figure \ref{fig:rec}.

\psfrag{rhoq}{$\rhos$} \psfrag{rhoc}{$\;\rhoc$}
\psfrag{sigc}{$\;\sigc$} \psfrag{rhohq}{$\rhohs$}
\psfrag{ccal}{$\Ccal$} \psfrag{ecal}{$ 
\Ecal$} \psfrag{rcal}{$\Rcal$}

{\small
\begin{figure}[h]
\centering
\epsfig{file=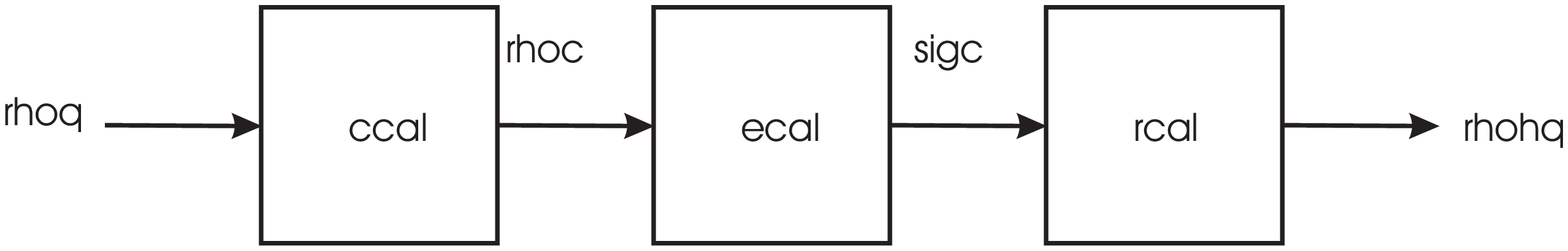,height=0.5in,width=3in}
\caption{Standard representation of error correction.}
\label{fig:rec}
\end{figure}
}

\noindent Here $\rhos$ is the $\ns\times\ns$ system state, $\rhoc$ is
the $\nc\times\nc$ encoded state, $\sigc$ is the $\nc\times\nc$
perturbed encoded state, and $\rhohs$ is the $\ns\times\ns$ recovered
system state. Using the OSR: \begin{equation}\label{rec osr} \rhohs =
\sum_{r,e,c} (R_r E_e C_c) \rhos (R_r E_eC_c)^\dag. \end{equation} The
encoding $\{C_c\}_{c=1}^\mc$ and recovery $\{R_r\}_{r=1}^\mr$
operation elements are rectangular matrices, respectively
$\nc\times\ns$ and $\ns\times\nc$, since they map between the system
Hilbert space of dimension $n_S$ and the system/ancillas Hilbert
space, the \emph{codespace}, of dimension $n_C$. The error operation
elements $\{E_e\}_{e=1}^\me$ are square $(\nc\times\nc)$ matrices, and
represent the effects of noise on the codespace. The number of
elements, $\mc,\me,\mr$ depend on the manner of implementation and
basis representation \cite{NielsenC:00}. More specifically, any OSR
can be equivalently expressed, and consequently physically
implemented, as a unitary with
ancilla states \cite[\S 8.23]{NielsenC:00}. An example of this representation of the standard
error correction model of Figure \ref{fig:rec} is shown in the block
diagram of Figure \ref{fig:urec}.

\psfrag{ccal}{$\Ccal$} \psfrag{ecal}{$\Ecal$} \psfrag{rcal}{$\Rcal$}
\psfrag{rhoc}{$\;\;\rho_{C}$} \psfrag{sigc}{$\;\sig_{C}$}
\psfrag{psira}{$\ket{0_{RA}}$} \psfrag{rhoq}{$\rho_S$} 
\psfrag{rhohq}{$\rho_R$} \psfrag{ox}{} 
\psfrag{rhosab}{$\rho_{SAB}$}

\psfrag{uc}{$U_C$} \psfrag{ue}{$U_E$} \psfrag{ur}{$U_R$} 
\psfrag{psica}{$\ket{0_{CA}}$} \psfrag{rhob}{\!\!$\rho_B$} 
\psfrag{rhor}{$\sig_R$} \psfrag{rhohq}{$\rhoh_S$}
\psfrag{anc}{\hspace{-2ex}Ancilla}

{\small
\begin{figure}[h]
\centering
\epsfig{file=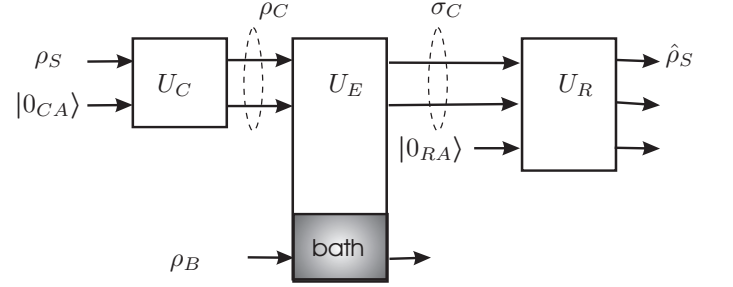,height=1.5in,width=3.75in}
\caption{System-ancilla-bath representation of standard
encoding-error-recovery model of error correction.}
\label{fig:urec}
\end{figure}
}

In this case the encoding operation $\Ccal$ is implemented by a
unitary operator $U_C$ acting on the (tensor) product of the system
state, $\rhos$, and the encoding
ancillas'
state, $\ket{0_{CA}}$, producing the encoded state
$\rhoc=U_C(\ket{0_{CA}}\bra{0_{CA}}\otimes\rhos)U_C^\adj $. 
(The tensor ordering is arbitrary, but once established must remain
fixed for consistency).
If the encoding
ancillas'
state has dimension $\nca$, then the resulting
codespace has dimension $\nc = \ns\nca$. If, as is customary, we take
$\ket{0_{CA}}$ as the $\nca$-column vector with a one in the first
element and zeros elsewhere (\emph{i.e.}, it is a tensor product of
$\log_2 n_{CA}$ encoding ancillas, each in the state $|0\rangle =
(1,0)^t$), then the OSR for $\Ccal$ has the \emph{single} ($\mc=1$)
$\nc\times\ns$ matrix element $C$ whose columns are the first $\ns$
columns of $U_C$, thus forming a set of orthonormal \emph{codewords},
\ie,
\begin{equation}
  \label{uc1}
  U_C = \left[C\ \cdots\right],\ \mbox{$C$ is
    $\nc\times\ns$}
\end{equation}

For the errors, $\Ecal$, the
ancillas' states are not implemented by
design, but rather, engendered by interaction with the \emph{bath}, a
term used to generically describe the physical environment.  The error
operation is thus equivalent to the unitary $U_E$ operating on the
tensor product of $\rhoc$, the encoded state, and $\rho_B$, the bath
state.  The number of bath states may be very large, in principal
infinite dimensional. However, it is always possible to represent
$\Ecal$ with a finite number of OSR elements with $\me \leq \nc^2$
\cite[Thm.8.3]{NielsenC:00}.

Finally, the recovery operation $\Rcal$ can be implemented via the
unitary $U_R$ operating on the (tensor) product of the perturbed
encoded state, $\sigc$, and the (additional) recovery ancillas'
state $\ket{0_{RA}}$. If $\ket{0_{RA}}$ is an $\nra$-column vector with a
one in the first element and zeros elsewhere, then the OSR
$\{R_r\}_{r=1}^\mr$ for $\Rcal$ has $\mr=\nca\nra$ elements which
consist of the first $\nc$ columns of $U_R$, \ie,
\begin{equation}\label{ur} U_R =
\left[R\ \cdots\right],\ R= \left[ \begin{array}{c} R_1\\
\vdots\\ R_{\mr} \end{array} \right],\ \mbox{$R_r$ is
$\ns\times\nc$}
\end{equation}

The model represented in Figures \ref{fig:rec} and \ref{fig:urec}
assumes that the encoding and recovery operations can be implemented
much faster than relevant time-scales associated with the bath. For a
detailed discussion of the validity of such a Markovian model see
\cite{AlickiLidarZanardi:05}.  Nevertheless, we will assume the model
of Figures \ref{fig:rec} and \ref{fig:urec} for the remainder of this
work, as complications associated with the bath being ``on'' during
encoding and recovery are likely to be dealt with via fault tolerance
methods \cite{Aliferis:05}, which require a base level of encoding of
the type we find here.

\begin{widetext}
Table \ref{definitions} provides definitions of some frequently used
symbols.

\begin{table}
\begin{center}
\begin{small}
\caption{Definitions of some frequently used symbols.}
\label{definitions}
\begin{tabular}{|l|l|}
\hline Symbol        &     ~~~~~~~~~~~~~~~~~~~~~~~~~~~~~~~~~~Definition              \\
\hline ~~$n_{S}$ & dimension of the system space           \\
\hline ~~$n_{CA}$ & dimension of the encoding ancillas space             \\
\hline ~~$n_C$ & dimension of the (system + encoding ancillas) space,
\emph{i.e.}, $n_C=n_S\times n_{CA}$      \\
\hline ~~$n_{RA}$ & dimension of the recovery ancillas space          \\
\hline ~~$m_E$ & number of operation elements for error map           \\
\hline ~~$m_{R}$ & number of operation elements for recovery map           \\
\hline \end{tabular}
\end{small}
\end{center}
\end{table}
\end{widetext}

\subsection{Performance measures}

Assume that we are given the OSR elements of the error channel
$\Ecal$. This could be obtained, for example, from the output of a
quantum process tomography experiment \cite{tom}. The error correction
objective considered here is to design the encoding $\Ccal$ and the
recovery $\Rcal$ so that, for a given error operation $\Ecal$, the map
$\rho_S \to \hat{\rho}_S$ is as close as possible to a desired
$\ns\times\ns$ unitary logic gate $\udes$.  Common measures of
performance between two quantum channels are typically based on
\emph{fidelity} or \emph{distance} \cite{NielsenC:00},
\cite{GilchristLN:04}, \cite{KretschmannW:04}, \cite{KosutGBR:06}.
Here we will use the \emph{channel fidelity} \cite{ReimpellW:05}
between the error correction operation $\Rcal\Ecal\Ccal$ and the
desired
operation
$\udes$:
\begin{equation}\label{favg} 
\favg = \frac{1}{n_s^2} \sum_{r,e,c}\ | \trace\ \udes^\dag R_r E_e
C_c |^2.
\end{equation}
where $0 \leq \favg \leq 1$ and from 
\cite{KnillL:97},
\cite[Thm.8.2]{NielsenC:00}, $\favg=1$ if and only if there are
constants $\delta_{rec}$ such that,
\begin{equation}\label{ind} R_r E_e C_c = \delta_{rec} \udes,\
\sum_{r,e,c} |\delta_{rec}|^2 = 1. \end{equation} This suggests the
\emph{indirect} measure of fidelity, the ``distance-like'' error
(using the Frobenius norm, $\normf{X}^2=\trace\ X^\dag X$),
\begin{equation}\label{dind} 
\begin{array}{rcl} 
\dind &=& \sum_{r,e,c}\ \normf{ R_r E_e C_c -
\delta_{rec} \udes }^2 \\
&=& \sum_c\ \normf{ R E(\Ie\otimes C_c)-\Delta_c\otimes \udes }^2
\end{array}
\end{equation}
where
\begin{equation}\label{delta}
\Delta_c\equiv[\delta_{rec}],~~~\dim\Delta_c=\mr\times\me,
\end{equation}
and $E$ is the $\nc\times\nc\me$ rectangular ``error matrix,''
\begin{equation}
  E=[E_1\ \cdots\ E_\me] ,
  \label{emat}
\end{equation}
$R$ is the $\mr\ns\times\nc$ matrix obtained by stacking the $\mr$
matrices $R_r$ as in \eqref{ur}, and $\Ie$ is the $\me\times\me$
identity. Hence, we have $\sum_c\trace\
\Delta_c^\dag\Delta_c=\sum_{r,e,c}|\delta_{rec}|^2=1$, and $R^\dag
R=\sum_r\ R_r^\dag R_r = \Ic $.

We show in Appendix \ref{sec:maxvr01} that there exists a recovery and
encoding pair, $\Rcal,\Ccal$, which achieves perfect error correction
(equivalently $\dind =0,\ \favg=1$), iff for
$c,c^{\prime}=1,\ldots,\mc$
\begin{equation}
  \label{klgen}
  (\Ie\otimes C_c^\adj) E^\adj E(\Ie \otimes C_{c'}) =
\Delta_c^\dag\Delta_{c'} \otimes \Is
\end{equation}
This is a generalization to non-unitary CP encoding of the
Knill-Laflamme condition for perfect error correction with unitary
encoding \cite{KnillL:97}. In this latter case, $\Ccal$ has only a
single $\nc\times\ns$ matrix element $C,~C^\dag C=\Is$, whose $\ns$
columns are the \emph{codewords}. As $\favg$ and $\dind$ are
explicitly dependent on the channel elements, they are convenient for
optimization. Consider then the following optimization problems.
\begin{equation}\label{favg opt}
\begin{array}{c}
\textit{ Direct Fidelity Maximization}\\
\begin{array}{ll}
\mathrm{maximize} & \favg(R,C) \\
\mathrm{subject~to} & R^\dag R = \Ic, ~~ C^\dag C = \Is
\end{array}
\end{array}
\end{equation}
\begin{equation}\label{dind opt}
\begin{array}{c}
\textit{Indirect Fidelity Maximization}\\
\begin{array}{ll} \mathrm{minimize} & \dind(R,C,\Delta_1,\ldots,\Delta_ {m_C} )
\\
\mathrm{subject~to} & R^\dag R = \Ic, \;\; C^\dag C = \Is,~~\sum_c
\normf{\Delta_c}^2=1 \end{array}
 \end{array}
\end{equation}
Here $C$ is the $\nc\times\ns$ matrix obtained in \eqref{uc1} . The
direct approach was used in
\cite{ReimpellW:05,FletcherSW:06,Fletcher:06,ReimpellWA:06,KosutL:06,
KosutAL:06}. As shown in Appendix \ref{sec:delta}, $\favg$ and $\dind$
are related as follows:
\beq[eq:fdrel]
\bea{rcl}
\favg(R,C) &=& \left( 1 - \hat{d}(R,C)/2\ns \right)^2
\\
\hat{d}(R,C) &=&
\min\set{\dind(R,C,\Del_1,\ldots,\Del_\mc}
	{\normf{\Delta_c}^2=1
,\; \forall c	  
	}
\eea
\eeq
This shows that minimizing the distance \eqref{dind opt} is equivalent
to maximizing fidelity \eqref{favg opt}.

\subsection{Robust error correction}

An important advantage of the method presented here is that unlike the
standard error correction model, it accounts for uncertainty in
knowledge of the channel. Such uncertainty may exist for many
reasons. For example, different runs of a tomography experiment can
yield different error channels $\{\Ecal_\alpha\}_{\alf=1}^\ell$. Or, a
physical model of the error channel might be generated by a
Hamiltonian $H(\theta)$ dependent upon an uncertain set of parameters
$\theta$. In any case, not accounting for the uncertainties typically
leads to non-robust error correction, in the sense that a small change
in the error model can lead to poor performance of the error
correction procedure. One way to account for these Hamiltonian
parametric uncertainties is to take a sample from the set of
Hamiltonians, say, $\{H(\th _\alpha)\}_{\alpha=1}^\ell$. Tracing out
the $m_E$ bath states will result in a \emph{set} of error
channels
$\{\Ecal_\alpha \}_{\alpha=1}^\ell$ where each error channel
has OSR elements $\{E_{\alpha,k} \}_{k=1}^\kappa$, where $\kappa$ is the
largest of the number of OSR elements in each sample. In those samples
with a smaller number we can set the corresponding OSR
elements to zero.

Two standard measures of robustness are the \emph{average-case} and
\emph{worst-case}. For the average-case, suppose that each OSR set
$\Ecal_\alf$ is known to occur with probability $p_\alf$. Then define
the \emph{average error channel} by the OSR,
\beq[eavg]
\Ecal^\av = \set{\sqrt{p_\alf}E_{\alf,k}}{\alf=1,\ldots,\ell,
k=1,\ldots,\kappa}
\eeq
The average error
channel in this form has $\kappa\ell$ OSR elements,
potentialy a very large number. However, this number is readily
reduced to no more than $\me=n_C^2$ using a singular value
decompostion \cite[Thm.8.3]{NielsenC:00}. Associated with $\Ecal^\av$
is the average channel fidelity,
\beq[fav]
\favg^\av = \sum_\alf p_\alf \favg_\alf
=\frac{1}{n_S^2}
\sum_{r,e',c}
|\trace\ R_r E^\av_{e'}C_c|^2
\eeq
where $E^\av_{e'},\ e'=1,\ldots,\ell\kappa$ are the OSR elements of
$\Ecal^\av$ in \eqref{eavg}.

For average-case robust error correction we replace $\favg$ in
\eqref{favg opt} with $\favg^\av$ in \eqref{fav}, and using the
relationship \refeq{fdrel}, replace $\dind$ in \eqref{dind opt} with,
\beq[dav opt]
\bea{rcl}
\dind^\av 
&=&  
\normf{R E^\av (\Ie \otimes C) - \Delta \otimes \udes}^2
\\
E^\av &=& [E^\av_1\ \cdots\ E^\av_{\ell\kappa}]
\eea
\eeq
A similar formulation exists for worst-case error correction which was
considered in \cite{KosutAL:06}; we do not consider it any further
here.
The remainder of the paper concentrates on the average-case objective
and development of the associated optimization algorithms.  The
examples presented in Sec.\ref{sec:examples} show that this approach
yields a high degree of robustness to uncertainty in the optimal
codes. 

We now discuss methods to approximately solve (obtain local solutions
to) the indirect optimization problem \eqref{dind opt}.

\section{Indirect Fidelity Maximization}

\label{sec:indirect}

We consider the encoding operator $\mathcal{C}$ as a unitary
operator acting on both the encoding ancillas and the input qubit.
Using the constraints in \eqref{dind opt}, we can express the
distance measure \eqref{dind} as
\begin{eqnarray}
&&\dind(R,C,\Delta)
= \normf{ R E (\Ie \otimes C) - \Delta \otimes
\udes}^2 
\label{dexpand}
\\
&&~=n_{S}+\text{Tr}~E(I_{E}\otimes CC^{\dagger })E^{\dagger }-2\text{Re}~\text{Tr}
RE(\Delta^{\dagger }\otimes CL_{s}^{\dagger })\nonumber
\end{eqnarray}
where $\Delta $ is the \emph{single} $m_{R}\times m_{E}$ matrix in
\eqref{delta} with $m_{R}=n_{CA}n_{RA}$ (note that in this case, since
there is only a single $\Delta_c$ matrix, we drop the subscript $c$).

\subsection{Optimal Recovery}

Since only the last term in \eqref{dexpand} depends on the recovery
matrix $R$, minimizing $\dind(R,C,\Delta)$ with respect to $R$ is
equivalent to maximizing the last term. In Appendix \ref{sec:maxvr01},
we show that this maximization results in
\begin{equation}\label{maxvr}
\underset{R^{\dagger }R=I_{C}}{\max }\text{Re}~ \text{Tr}~ RE(\Delta
^{\dagger }\otimes CL_{s}^{\dagger})=\text{Tr}~\sqrt{E(\Gamma \otimes
CC^{\dagger })E^{\dagger }},
\end{equation}
where the $m_{E}\times m_{E}$ matrix $\Gamma$ is defined as,
\begin{equation}
\Gamma =\Delta ^{\dagger }\Delta,
\label{gamma}
\end{equation}
and the associated
$n_{C}n_{RA}\times n_{C}$ optimal recovery matrix
is,
\begin{equation}
\label{ropt ind}
R=[v_{1}\ ...\ v_{n_{C}}][u_{1}\ ...\ u_{n_{C}}]^{\dagger}
\end{equation}
where $v_{i},\ u_{i},\ i=1,\ldots,n_C$ are, respectively, the right
and left singular vectors in the singular value decomposition of the
$n_{C}\times n_{C}n_{RA}$ matrix $E(\Delta ^{\dagger }\otimes
CL_{S}^{\dagger})$, with the singular values, as usual, in descending
order.  Thus, to obtain the optimal recovery, we need first to find
$\Gamma$ which maximizes \eqref{maxvr} -- this is equivalent to
minimizing $\dind$ over $R$. Following this we need to determine
$\Delta$ satisfying \eqref{gamma}.

To find $\Gamma$, observe that $\Gamma\geq 0$ by definition
\eqref{gamma}, and the constraint $\normf{\Delta}=1$ from \eqref{dind
opt} is equivalent to $\mathrm{Tr}\ \Gamma=1$. Hence, optimal recovery
can be obtained by first solving for $\Gamma $ from,
\begin{equation}
\label{optgamma} 
\begin{array}{ll} 
\mathrm{maximize} 
&
\mathrm{Tr}\sqrt{E(\Gamma \otimes CC^{\dag})E^{\dag}}
\\
\mathrm{subject~to} 
& 
\Gamma \geq 0,~ \text{Tr}~\Gamma =1
\end{array} \end{equation}
In Appendix \ref{sec:gamopt} it is shown that the optimal $\Gamma $ is
the solution of an equivalent SDP.

The next step is to use \eqref{gamma} to obtain $\Delta $ from
$\Gamma$.  The following choice adheres to the given dimensions:
\beq[eq:gam2del]
\bea{ccl}
\left(
\bea{c}
\nca \leq \me
\\
\nra\nca=\me
\eea
\right)
&
\Rightarrow
&
\left\{
\bea{l}
\Del = \sqrt{\Gam}
\\
\mbox{$R$ is tall ($\ns\me\times\nc$)}
\eea
\right.
\\
\left(
\bea{c}
\nca>\me
\\
\nra=1
\eea
\right)
&
\Rightarrow
&
\left\{
\bea{l}
\Del=\left[
\bea{c} \sqrt{\Gam} \\ 0_{\nca-\me\times\me} \eea
\right]
\\
\mbox{$R$ is unitary ($\nc\times\nc$)}
\eea
\right.
\eea
\eeq
Clearly the choice of $\Del$ is not unique. In fact, the result does
not change if $\Delta$ is multiplied by a unitary, \emph{i.e.},
$\Delta\rightarrow U\Delta$. This is exactly the unitary freedom in
choosing the OSR elements \cite{NielsenC:00}. Interestingly, however,
from many numerical calculations we observe that the
following holds:
\begin{equation}
\begin{cases}
\text{rank}(\Gamma)=n_{CA}~~ \text{if}~~ n_{CA}\leq m_{E}\\
\text{rank}(\Gamma)=m_{E}~~ \text{if}~~ n_{CA} >m_{E}.
\end{cases}
\label{eq:rank gam}
\end{equation}
Since the $m_{E}\times m_{E}$ matrix $\Gamma$ is Hermitian $( =
\Delta^{\dag} \Delta$), and $\Del$ is $\mr\times\me$ with $\mr=\nca\nra$,
it follows that if \refeq{rank gam} is true then,
\begin{eqnarray}
    n_{RA}=1.
\end{eqnarray}
If, in the optimized error correction, \emph{the dimension of the
recovery ancillas space is one, then the optimal recovery matrix $R$
is always a unitary -- recovery ancillas are redundant in maximizing
the fidelity.} Note that we started with a generic $n_{RA}$ parameter,
and the properties of the optimal solution led us to the above
conclusion.  Although we do not have a rigorous proof that the
recovery ancillas are redundant, a compelling heuristic argument is
offered in Section \ref{sec:ancilla} along with supporting numerical
results.

\subsection{Optimal Encoding}

For a given $R$ and $\Delta $, the optimal encoding $C$ can be found
by solving \eqref{dind opt} for $C$, that is,
\begin{equation}
  \label{copt}
\begin{array}{ll} \mathrm{minimize} & 
\dind(R,C,\Del) = 
\normf{R E (\Ie \otimes C) - \Delta \otimes \udes}^2
  \\ \mathrm{subject~to} & C^\adj C = \Is
\end{array}
\end{equation} 
As shown in Appendix \ref{sec:copt}, the optimal encoding $C$ is given
by,
\begin{equation}
\label{copt svd1}
C = UV^\dag
\end{equation}
where $(U,V)$ are obtained from the SVD,
\begin{equation}
\label{copt svd2}
\Cb = \sum_{r,e} \delta_{re}(R_rE_e)^\dag L_S
= USV^\dag \nonumber
\end{equation}
with $U$ an $\nc\times\ns$ matrix with orthonormal columns, \ie,
$U^\dag U=\Is$, $V$ an $\ns\times\ns$ unitary, and $S$ a diagonal
matrix of the $\ns$ singular values.  The matrix $\Cb$ is the
\emph{unconstrained} (least-squares) solution to \eqref{copt}, \ie,
$\min_C\dind$.

The left-hand column of Table \ref{tab:alg}, labeled \algsdp,
summarizes the preceding method for recovery and encoding
optimization. For optimal recovery alone, solve \eqref{optgamma} for
$\Gamma$, then determine $\Del$ via \refeq{gam2del}, and finally $R$
from \eqref{ropt ind}. For optimal encoding alone, solve \eqref{copt}
for $C$. To find a combined optimal encoding and recovery repeat steps
1 and 2 in Table \ref{tab:alg} until $\dind$ stops decreasing. (By
virtue of \refeq{fdrel}, fidelity increases in every step). Since in
each step the distance measure, $\dind$, can only decrease, never
increase, the converged solution to the combined optimization is only
guaranteed to be a local optimal solution to \eqref{dind opt}.

\subsection{Alternative iterative algorithm for recovery optimization}

An alternative to the above optimal recovery procedure (Step 1 in
\algsdp of Table \ref{tab:alg}) is to iterate between solving
\eqref{dind opt} \emph{directly} by minimizing over $\Del$ and then
using \eqref{ropt ind} to find $R$. Specifically, for a given $R$ and
$C$, Step 2a in \algcls of Table \ref{tab:alg} requires solving
the following constrained least-squares problem for $\Del$:
\begin{equation}
  \label{delopt}
\begin{array}{ll} \mathrm{minimize} & 
\dind(R,C,\Del) = \normf{R E (\Ie \otimes C) - \Delta \otimes \udes}^2
  \\ \mathrm{subject~to} & \normf{\Del}=1
\end{array}
\end{equation} 
As shown in Appendix \ref{sec:delta}, the solution is,
\begin{equation}\label{delta cls}
\begin{array}{rcl}
\Delta &=& \bar{\Delta} / \normf{\bar{\Delta}}
\\
\left(\bar{\Delta}\right)_{re} &=& \trace(R_rE_e C\udes^\dag)/\ns,
\end{array}
\end{equation}
where $\bar{\Delta}$ is the unconstrained (least squares) solution to
$\min_{\Delta}\dind$. This solution is then used in \eqref{ropt ind}
to find $R$ (Step 1b), then back to \eqref{delta cls} (Step 1a), and
so on until $\dind$ stops decreasing (Step 1c).
\begin{widetext}
\begin{em}
\bc
\begin{table}[hbt]
\caption{Iterative Algorithms for Optimal QEC}
\label{tab:alg}
\btab{|c|c|}
\hline
{\bf \algsdp} & {\bf \algcls}
\\
\hline
\parbox{2.5in}{
\begin{description}
\item {\bf Initialize} $R$ and $C$
\item {\bf Repeat}
	\ben
\item {\bf Optimal recovery}
		\ben
\item solve \eqref{optgamma} for $\Gam$
\item $\Gam\to\Del$ via \refeq{gam2del}
\item $\Del\to R$ via \eqref{ropt ind}
		\een
\item {\bf Optimal encoding}
		\ben
\item solve \eqref{copt} for $C$
		\een
	\een
\item {\bf Until} $\dind$ stops deceasing
\end{description}
} 
&
\parbox{3.5in}{
\begin{description}
\item {\bf Initialize} $R$ and $C$
\item {\bf Repeat}
	\ben
\item {\bf Optimal recovery} -- Repeat a-c 
		\ben
\item solve \eqref{delopt} for $\Del$
\item $\Del\to R$ via \eqref{ropt ind}
\item Until $\dind$ stops decreasing
		\een
\item {\bf Optimal encoding}
		\ben
\item solve \eqref{copt} for $C$
		\een
	\een
\item {\bf Until} $\dind$ stops deceasing
\end{description}
} 
\\{}&{}\\
\hline
\etab
\end{table}
\ec
\end{em}
\end{widetext}
The difference between the two algorithms is in computing the optimal
recovery (Steps 1).  In Step 1 of \algsdp, no iterations are required;
the optimal recovery is achieved 
by solving the SDP \eqref{optgamma}.  
For Step 1 of \algcls, an optimal recovery is the result of some
number of iterations
involving the constrained least-squares problem \eqref{delopt}.
Although at present a proof is not available, in every case we have
tried the optimal fidelity in both recovery algorithms converges to
the same result. Additionaly, the total CPU-time in MATLAB to compute
the optimal recovery in \algcls\ (including the iterations) is
significantly less than the CPU-time for the recovery step in \algsdp\
using YALMIP \cite{Yalmip:04} to call the solver SDPT3 \cite{Sdpt3}.

\section{Dimension of the Recovery Ancillas space}
\label{sec:ancilla}

In our formalism, the dimension of the Recovery ancillas'
space, \emph{i.e.}, the required number of recovery ancilla qubits, is
determined by the rank of the $m_{E}\times m_{E}$ matrix $\Gamma$. 

\subsection{Rank minimization of $\Gamma$}

In this section, we study the rank
of $\Gamma$ through a heuristic argument by noting the similarity
between our problem and the so called
``Rank Minimization Problem'' (RMP) \cite{boyd:96}:
\begin{equation}\label{rankmin}
\begin{array}{ll} \mathrm{minimize} & \mathrm{rank}~(X)
\\
\mathrm{subject~to} & X\in \Xcal \end{array} 
\end{equation}
The matrix $X$ is the optimization variable and $\Xcal$ is a convex
set denoting the constraints.

Although several special cases of the RMP have well-known solutions,
in general the RMP is known to be computationally
intractable. However, there are a number of heuristic approaches to
solving this problem. Restate \eqref{optgamma} as follows,
\begin{equation}
  \label{restated}
  \begin{array}{ll} \mathrm{minimize} & \mathrm{Tr}~(\Gamma)
\\
\mathrm{subject~to} & \Gamma\geq
0,~~\mathrm{Tr}\sqrt{E(\Gamma\otimes CC^{\dag})E^{\dag}}\geq
\text{const.}
\end{array}
\end{equation}
where the constant is the maximum which arose in~\eqref{optgamma}.  A
well known heuristic for RMP when $X$ is positive semidefinite
\cite{fazel:02,fazel:04,mesbahi:97} is to replace the rank objective
with $\text{Tr}~[X]$ and solve,
\begin{equation}
  \label{traceh}
  \begin{array}{ll}
    \mathrm{minimize} & \mathrm{Tr}~[X]
\\
\mathrm{subject~to} & X\in
\Xcal,~X\geq 0
  \end{array}
\end{equation}
By comparing~\eqref{restated} with~\eqref{traceh}, we can view our
problem in~\eqref{optgamma} as an RMP that minimizes the rank of
$\Gamma$. Thus, the rank of the optimal $\Gamma$ is the smallest
possible consistent with not changing the rank of our objective
matrix, $\sqrt{E(\Gamma\otimes CC^{\dag})E^{\dag}}$. Noting that
rank$(CC^{\dagger})=n_S$ and rank$(E)=n_C$ and with a straightforward
linear algebra analysis we find that this property holds if
\begin{equation}
\begin{cases}
\text{rank}(\Gamma)\geq n_{CA}~~ \text{if}~~ n_{CA}\leq m_{E}\\
\text{rank}(\Gamma)=m_{E}~~ \text{if}~~ n_{CA} >m_{E}.
\end{cases}
\end{equation}
That is, in the first case, if $\text{rank}(\Gamma)<
n_{CA},~\text{rank}(\sqrt{E(\Gamma\otimes CC^{\dag})E^{\dag}})$
decreases by decreasing the rank of $\Gamma$. But if
$\text{rank}(\Gamma)\geq n_{CA}$, $\text{rank}(\sqrt{E(\Gamma\otimes
CC^{\dag})E^{\dag}})=n_{C}$, and it does not depend on
$\text{rank}(\Gamma)$. In the second case, $\Gamma$ should be full
rank. Therefore the rank of the optimal $\Gamma$ is
\begin{equation}
\begin{cases}
\text{rank}(\Gamma_{\mathrm{opt}})=n_{CA}~~ \text{if}~~ n_{CA}\leq m_{E}\\
\text{rank}(\Gamma_{\mathrm{opt}})=m_{E}~~ \text{if}~~ n_{CA} >m_{E},
\end{cases}
\end{equation}
which agrees with \refeq{rank gam}.
Note that the same argument also applies in the average case
\eqref{dav opt} with $E$ replaced by $E^\av$.

\subsection{Numerical result for randomly generated error maps}

Here, we examine the result above for randomly generated error
maps. Namely, we find the rank of the optimal $\Gamma$ for each random
map by applying the indirect optimization method. The error map is
modeled as shown in Fig.~\ref{fig:urec} as a unitary $U_{E}$ acting on
the joint codespace-bath Hilbert space. The unitary $U_{E}$ arises
from a randomly selected $m_{E}n_{C}\times m_{E}n_{C}$
time-independent Hamiltonian $H_{E}$,
\ie, $U_E = e^{-i t H_E}$ (we work in units where $\hbar = 1$).
The unitary evolution operator generated by this Hamiltonian
at time $t=1$ is
\begin{eqnarray}
    U_{E}=\exp(-iH_{E})=\left[\begin{array}{c} E_1~~\ldots\\
E_2~~\ldots\\
\vdots\\
E_{m_{E}}~~\ldots\end{array}\right] \label{rmap}
\end{eqnarray}
That is, we pick the first $n_{C}$ columns of the matrix $U_{E}$.
Here, $E_1\ldots E_{m}$ are the $n_{C}\times n_{C}$ OSR elements of
the error operation, and from \eqref{emat}, $E=[E_1\ \cdots\ E_\me]$.

\begin{figure}[h]
\begin{center}
\btab{c} \epsfig{file=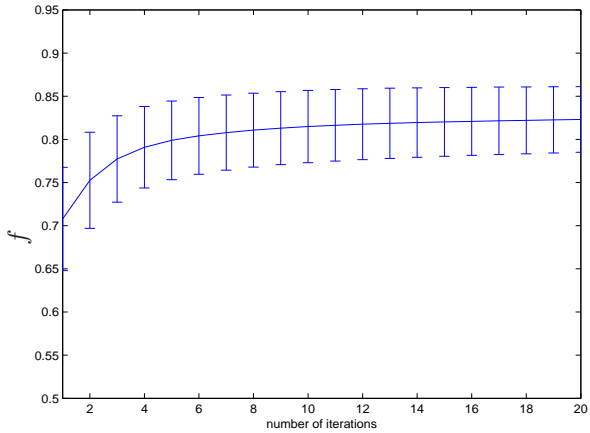,width=3.5in,height=2.5in} \etab
 \end{center}
\caption{Channel fidelity $\favg$ for random error maps on two-qubit
  codes.  }
\label{meanf}
\end{figure}

\begin{figure}[h]
\begin{center}
\btab{c} \epsfig{file=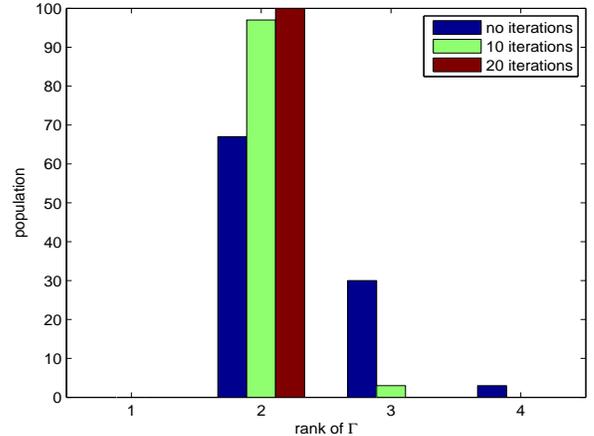,width=3.5in,height=2.5in} \etab
 \end{center}
\caption{Rank of the optimal $\Gamma$ for random error maps on
  two-qubit codes.
  \label{hist}}
\end{figure}

Figure~\ref{meanf} presents the channel fidelity \emph{vs.} the number
of iterations in Algorithm $1$ for 100 random error maps. In this
experiment, the
system is a single qubit and one qubit is used
as an encoding ancilla, \emph{i.e.}, $n_{S}=2$, $n_{CA}=2$. Each error
map has 4 OSR elements, \emph{i.e.}, $m_E = 4$, and is generated using
a $16\times 16$ random Hamiltonian matrix according to \eqref{rmap}.
Therefore, the matrix $\Gamma$ in \eqref{optgamma} is $4\times 4$.
Figure~\ref{hist} shows the histogram of the rank of $\Gamma$
\emph{vs.}  the number of iterations. This histogram indicates that
after $20$ iterations in the optimization algorithm, the rank of
$\Gamma$ is always two, which is equal to $n_{CA}$. In fact, those
$\Gamma$ that are not rank 2 after 10 iterations are associated to the
error maps with lower rate of fidelity convergence.

\psfrag{average fidelity}{$\favg$}
\psfrag{fidelity}{\small $\favg$}
\psfrag{p}{\small $p$}
\psfrag{probability}{\small $p$}
\psfrag{probability of error}{\small $p$}

\begin{figure}[h]
\begin{center}
\btab{c} \epsfig{file=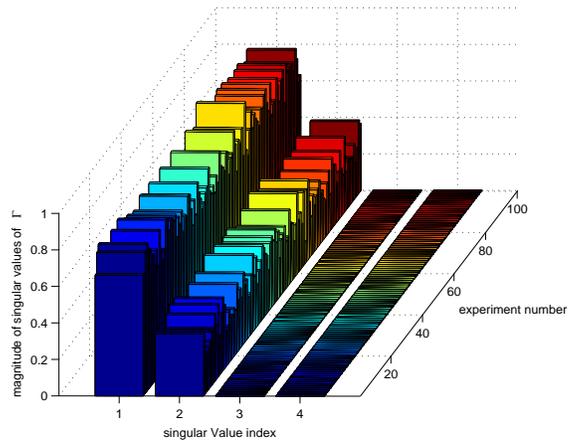,width=3.5in,height=2.5in} \etab
\caption{
Singular values of the optimal $\Gamma$ for random error maps on
two-qubit codes. For all cases tested only two of the singular values
are significantly different from zero, meaning that the rank of the
$\Gamma$ matrices is $2$.
   \label{singularvalue}}
\end{center}
\end{figure}

Figure~\ref{singularvalue}, which shows the singular values of the
same $\Gamma$ matrices, is included for comparison of the magnitude of
the singular values. In all cases, the nonzero singular values are of
the order of $10^{-1}$. The numerical precision of all the results is
$10^{-8}$.
We repeated the experiment for more than 1000 random maps with
different dimensions (only 100 are shown), and the result holds for
all of them. Namely, after sufficiently many iterations in Algorithm
$1$, \emph{the rank of the optimal $\Gamma$ is the same as the
dimension of the encoding ancillas space, i.e., rank
$(\Gamma_{opt})=n_{CA}$.}

\section{Examples}
\label{sec:examples}

We now apply the methods developed above to the goal of preserving a
single qubit ($\ns=2$) using a ${q_C}$-qubit ($\nc=2^ {q_{C}} $)
codespace. In these examples, the error channel
$\Ecal$ consists of single-qubit errors occurring \emph{independently}
on all qubits with probability $p$. We examine two cases of bit-flip
and bit-phase-flip errors.

\subsection{3-qubit bit-flip errors}

In this example, we consider the independently occurring bit-flip
error as the noise channel, where the bit-flip operator is
$X=${\tiny{$\left[\begin{array}{cc} \text{\small{0}} & \text{\small{1}}
\\
\text{\small{1}} & \text{\small{0}} \end{array}\right]$}}.
We used $q_C=2$ encoding ancilla qubits. There are $2^3=8$ OSR error elements for 3-qubit encoding:
\begin{equation}
  \label{bitflip3}
  \begin{array}{rcl} \seq{E_i}_{i=1}^8 &=& A_{i_1}\otimes
A_{i_2}\otimes A_{i_3}, \;\; i_1,i_2,i_3 \in \seq{1,2} \\
A_1 &=& \sqrt{(1-p)}\ I \;\; \mathrm{(no~error)} \\
A_2 &=& \sqrt{p}\ X \;\; \mathrm{(bit-flip~error)}
  \end{array}
\end{equation}
\begin{figure}[tbp]
\begin{center}
\btab{c} \epsfig{file=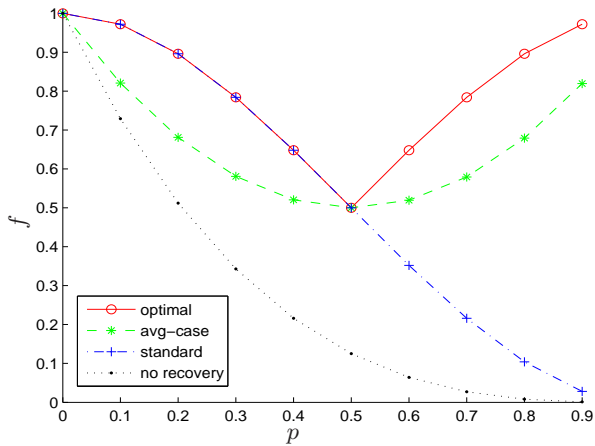,width=3.5in,height=2.5in} \etab
\caption{ Channel fidelity $\favg$ \emph{vs.}  bit-flip
probability $p$ for 3-qubit encoding.}
\label{fig:bitflip00}
\end{center}
\end{figure}

Figure \ref{fig:bitflip00} shows $\favg$ \emph{vs.}  bit-flip
probability $p$ in the range $p \leq 0.9$ for the standard 3-qubit
code, optimal recovery at each $p$, average-case recovery over the $p$
range, and no recovery.
For the average case, we computed an optimized encoding and recovery
for the single channel obtained by averaging over the error channels
corresponding to $p=0, 0.1, ..., 0.9$ as defined in \eqref{dav
opt}. We then applied this encoding and recovery to each of these $10$
channels, thus producing the $10$ fidelity values shown.
Note that the optimal recovery can be achieved equivalently by either
the constrained least squares method (\algcls) or the convex
optimization method (\algsdp). Interestingly, the standard 3-qubit
code not only provides optimal recovery for the range $p\leq 0.5$, it
is optimal for \emph{both} recovery and encoding in this range. For
$p>0.5$ the standard code is clearly no longer optimal. Only in this
range does the optimal recovery outperform the standard code, a
phenomenon similar to what was reported for amplitude-damping errors
in \cite{ReimpellW:05}.
Analysis of our optimal encoding recovery results reveals the
following simple picture. The optimal code is the standard $3$-qubit
code for the entire $p$ range, \ie, $\ket{\bar{0}}=\ket{000}$ and
$\ket{\bar{1}}=\ket{111}$. The optimal recovery is the standard
recovery \cite{NielsenC:00} in the range $0\le p \le 0.5$. In the
range $0.5 \le p \le 0.8$ the optimal recovery is a bit-flip on all
qubits followed by the standard recovery.

\begin{figure}[tbp]
\begin{center}
\btab{c} \epsfig{file=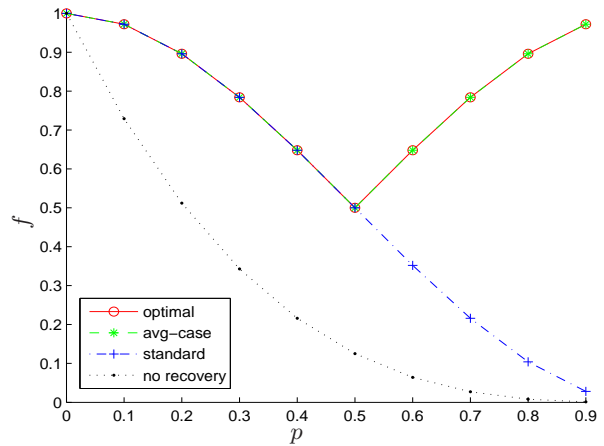,
width=3.5in,height=2.5in} \etab
\caption{Channel fidelity $\favg$ \emph{vs.} bit-flip probability $p$
for 3-qubit encoding in two ranges: $p<0.5$ and $0.5 < p \leq 0.9$.}
\label{fig:bitflip01}
\end{center}
\end{figure}

Figure \ref{fig:bitflip01} shows channel fidelity $\favg$ in two
ranges: $p<0.5$ and $0.5<p\leq 0.9$. Unlike the previous case, here we
compute the optimization twice, once for each range.
For the average case, we computed an optimized encoding and recovery
for the single channel obtained by averaging over the error channels
corresponding to $p=0, 0.1, ..., 0.4$. We then applied this encoding
and recovery to each of these $5$ channels, thus producing the
$5$ fidelity values shown in the range $0\leq p \leq 0.5$. We then
repeated this procedure for $p=0.5, 0.6, ..., 0.9$.
For $p < 0.5$, the standard, optimal, average-case, 
all coincide. For $p>0.5$, the optimal and
average-case codes coincide and divert again from the standard.
The optimal encoding and recovery are the same as in Figure
\ref{fig:bitflip00}, \ie, the standard $3$-qubit code, with standard
recovery in the range $0\le p \le 0.5$, and bit-flips preceeding
standard recovery in the range $0.5 \le p \le 0.9$.
We conclude from the examples in Figures \ref{fig:bitflip00} and
\ref{fig:bitflip01} that optimal encoding and recovery has no advantage over
standard encoding and recovery 
for low bit-flip probabilities
($p<0.5$), and thus increasing the codespace would be required to
improve fidelity. For large errors ($p>0.5$), optimization is more
effective in that it identifies an optimal recovery.
In both cases the achieved optimal fidelity is
independent of the number of recovery ancillas used, hence in all
examples shown in Figures \ref{fig:bitflip00} and
\ref{fig:bitflip01} there are no 
additional recovery ancillas required.
It is striking that the average case fidelity matches the optimal in
Figure \ref{fig:bitflip01}, but not in Figure
\ref{fig:bitflip00}. This is entirely due to the range of $p$ values
over which the average is performed.
The lesson is that the more information is available about the noise
channel, the more robust the encoding and recovery will be: in Figure
\ref{fig:bitflip01} we know that the probability is in the range
$[0,0.5]$ or $[0.5,0.9]$, while in Figure \ref{fig:bitflip00} we only
know that it is in the range $[0,0.9]$.  Absent such information,
robustness may still be attainable by experimenting with tuning the
encoding and recovery over a range of channels.

\subsection{Bit-Phase flip error}

In this example, the noise channel consists of
bit-phase flip errors
$Y=$ {\tiny{$\left[\begin{array}{cc}
\text{\small{0}} & -\textit{\small{i}} \\ \textit{\small{i}} &
\text{\small{0}} \end{array}\right]$}} occurring independently with
probability $p$. We do not allow for more than three to occur
simultaneously
(\ie, we consider weight-$3$ errors).
We examine two cases: 1. Considering a fixed number of
encoding ancillas, we compare the fidelity using different numbers of
recovery ancillas. 2. We fix the total number of available ancilla
qubits, and compare the fidelity for various distributions of encoding
and recovery ancillas.

\subsubsection{5-qubit bit-phase flip error}

In this example, the bit-phase flip errors occur independently on
the input qubit and 4 ancillas. There are 26 error OSR elements: 1
for no error, 5 for a single error, 10 for double errors, and 10 for
triple errors. Thus the matrix $\Gamma$ in \eqref{optgamma} is
$26\times 26$ and the rank of $\Gamma_{\rm opt}$ is equal to
$n_{CA}=16$,
meaning that the optimal distribution of ancillas is having all four
in the encoding block and none in the recovery block.

\begin{figure}[tbp]
\begin{center}
\btab{c} \epsfig{file=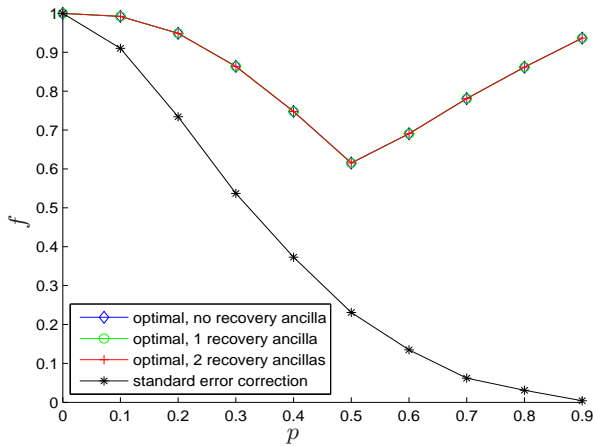,
width=3.5in,height=2.5in} \etab
\caption{ Channel fidelity $\favg$ \emph{vs.} bit-phase flip
 probability $p$ for 3 qubit code and 0, 1, or 2 recovery ancillas, 
with optimal encoding and recovery.}
\label{fig:bit-phase0}
\end{center}
\end{figure}

Figure \ref{fig:bit-phase0} shows $f$ \emph{vs.} bit-phase flip
error probability $p$ for the optimal encoding/recovery in the case of
zero, one and two recovery ancillas. The result shows that all cases yield the
same fidelity. Therefore, the fidelity of the system is independent of
the number of recovery ancillas.

\subsubsection{bit-phase flip errors with a fixed number of ancillas}

\begin{figure}[tbp]
\begin{center}
\btab{c} \epsfig{file=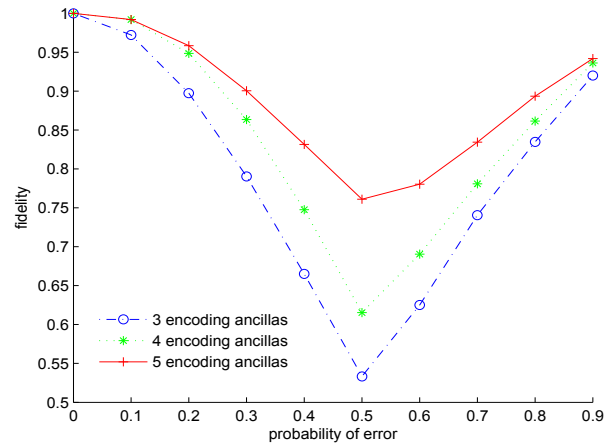,
width=3.5in,height=2.5in} \etab
\caption{ Channel fidelity $\favg$ \emph{vs.} bit-phase flip
  probability $p$ with a fixed total of 6 ancillas, and optimal
  encoding and recovery.}
\label{fig:bit-phase1}
\end{center}
\end{figure}

In this example, we consider six ancilla qubits that can be used
either in the encoding block or in the recovery block. We compare the
fidelity for the following distributions: four encoding ancillas and
two recovery ancillas, five encoding ancillas and one recovery
ancilla, and six encoding ancillas with no recovery ancilla. Figure
\ref{fig:bit-phase1} shows that the channel fidelity increases
significantly by using the ancillas in the encoding instead of the
recovery.
Thus the most efficient use of ancillas is achieved when they are all
used for encoding.

\section{Conclusion}

\label{sec:conclusions}

We have presented an optimization approach to quantum error correction
that yields codes which achieve robust performance, when tuned to a
specific noise channel. An important aspect of developing optimal
codes which are tuned to a class of errors, or are robust over a range
of errors, is that the optimized performance levels may be sufficient
for the intended purposes.
\emph{Hence, no further increases in codespace dimension may be necessary.}
This cannot be known without performing the optimization.

We also showed that the fidelity of such a system is independent of
the number of the recovery ancillas. This is entirely due to the
structure of the error correction optimization problem, for which we
found that a unitary recovery operator maximizes the fidelity of the
system. However, the fidelity increases significantly by increasing
the dimension of the encoding ancillas space. Therefore, in the
optimal quantum error correction scheme, one should use all the
available ancilla qubits in the encoding block.

Although not further developed here, the resulting codes, unlike
standard codes, have support over all basis states. Some of the
recovery structure is revealed via the indirect approach. This in turn
leads to a method for approximating optimal recovery involving only a
singular value decomposition, making it potentially useful in
evaluating very large blocks of encoding to see if further performance
improvement is possible.

We stress that there is an important difference between the standard
error correction schemes
\cite{Shor:95,Gott:96,Steane:96,Laflamme:96,KnillL:97,NielsenC:00} and
the approach presented here. While in the standard case only the class
of errors should be known, in our method the exact form of the noise
map is required for optimization.  In general, the noise map can be
identified using quantum process tomography \cite{tom}. In most cases
this extra knowledge is equivalent to identifying the probability of
the error, which can also be found using our method.  In order to
identify the probability in a particular error model, one should
calibrate the fidelity of the system using a fixed pair of recovery
and encoding operators. Once the relation between the fidelity
associated to this pair and the error probability is known, a
measurement of the fidelity yields the probability.

It thus appears that the effectiveness of optimization is dependent
upon the \emph{structure} of the error operation, a result seemingly
heralded by Feynman \cite{Feynman:QC1}:

{\small \bquote
``In a machine such as this there are very many other problems due to
  {imperfections}. \ldots\ there may be small terms in the Hamiltonian
  besides 
the ones we've written. \dots\ At least some of these problems can be
remedied in the usual way by techniques such as {error correcting codes}
\ldots\ But until we find a specific implementation for this computer, I do
not know how to proceed to analyze these effects. However, it appears that
they would be very important in practice. This computer seems to be very
delicate and these imperfections may produce considerable havoc.'' \equote
}

Determining the ``specific implementation'' is currently an on-going
research effort. Analyzing the ``effects'' however will undoubtedly be
accomplished by a combination of physical modeling and/or system
identification (\eg, process tomography and parameter
estimation). This leads to an intriguing prospect: to integrate the
results found here within a complete ``black-box'' error correction
scheme, that takes quantum state or process tomography as input and
iterates until it finds an optimal error correcting encoding and
recovery.

\section*{Acknowledgments}

Funded under the DARPA QuIST Program and (to D.A.L.) NSF CCF-0726439 and
NSF PHY-0802678.


\begin{thebibliography}{99}
\bibitem{Shor:95} P.~W. Shor, Phys. Rev. A \textbf{52}, R2493 (1995).

\bibitem{Gott:96} D. Gottesman, Phys. Rev. A \textbf{54}, 1862 (1996).

\bibitem{Steane:96} A.~M. Steane, Phys. Rev. Lett. \textbf{77}, 793 (1996).

\bibitem{Laflamme:96} {R. Laflamme, C. Miquel, J.P. Paz and W.H. Zurek},
Phys. Rev. Lett. \textbf{77},  198 (1996).

\bibitem{KnillL:97} E. Knill and R. Laflamme, Phys. Rev. A \textbf{55}, 900
(1997).

\bibitem{NielsenC:00} {M.A. Nielsen and I.L. Chuang}, \emph{Quantum
Computation and Quantum  Information} ({Cambridge University Press},
Cambridge, UK, 2000).

\bibitem{ReimpellW:05} {M. Reimpell and R. F. Werner}, Phys. Rev. Lett.
\textbf{94}, 080501 (2005).

\bibitem{YamamotoHT:05} N. Yamamoto, S. Hara, and K. Tsumara, Phys. Rev. A
\textbf{71}, 022322 (2005).

\bibitem{KosutL:06} {R. L. Kosut and D. A. Lidar}, eprint quant-ph/0606078.

\bibitem{KosutAL:06} R.L. Kosut, A. Shabani, and D.A. Lidar, Phys. Rev. Lett.
  \textbf{100}, 020502 (2008).

\bibitem{tom}  J. F. Poyatos, J. I. Cirac, and P. Zoller,
    Phys. Rev. Lett. {\bf 78}, 390 (1997); I. L. Chuang and
    M. A. Nielsen, J. Mod. Opt. {\bf 44}, 2455 (1997);
    G. M. D'Ariano and P. Lo Presti, Phys. Rev. Lett. {\bf 86}, 4195
    (2001); 
R. L. Kosut, I. A. Walmsley, and H. Rabitz,
eprint quant-ph/0411093 (2004);
M. Mohseni and D.A. Lidar, Phys. Rev. Lett. {\bf 97}, 170501 (2006).

\bibitem{BoydV:04} S. Boyd and L. Vandenberghe, \emph{Convex Optimization}
(Cambridge University  Press, Cambridge, UK, 2004).

\bibitem{Gaitan:book} {F. Gaitan}, \emph{Quantum Error Correction and
  Fault Tolerant Quantum Computing} (CRC Press, Boca Raton, 2008).

\bibitem{ShabaniLidar:08} {A. Shabani and D.A. Lidar}, 
eprint arXiv.org:0808.0175.
  
\bibitem{FletcherSW:06} A.~S. Fletcher, P.~W. Shor, and M.~Z. Win, Phys.
Rev. A \textbf{75}, 012338 (2007).

\bibitem{Fletcher:06} A.~S. Fletcher, Ph.D. Dissertation, eprint
arXiv:0706.3400.

\bibitem{ReimpellWA:06} {M. Reimpell, R.F. Werner, and K. Audenaert}, eprint
quant-ph/0606059.

\bibitem{AlickiLidarZanardi:05} {R. Alicki, D.A. Lidar, and P. Zanardi},
Phys. Rev. A \textbf{73}, 052311 (2006).

\bibitem{Aliferis:05} {P. Aliferis, D. Gottesman, and J. Preskill}, Quantum
Inf. Comput. \textbf{6}, 97 (2006).

\bibitem{GilchristLN:04} A.~Gilchrist, N.~K. Langford, and M.~A. Nielsen,
Phys. Rev. A \textbf{71}, 062310 (2005).

\bibitem{KretschmannW:04} D.~Kretschmann and R.~F. Werner, New J. of Phys.
\textbf{6}, 26 (2004).

\bibitem{KosutGBR:06} R.~L. Kosut, M.~Grace, C.~Brif, and H.~Rabitz, eprint
quant-ph/0606064.

\bibitem{horn:91}R.~Horn and C.~Johnson. \emph{Topics in Matrix Analysis} (Cambridge
University Press, Cambridge, 1991).

\bibitem{boyd:96}L.~Vandenberghe and S.~Boyd, SIAM Review {\bf 38}, 49 (1996).

\bibitem{fazel:02}M.~Fazel, "Matrix rank minimization with applications", PhD thesis,
Dept. of Electrical Engineering, Stanford University (2002).

\bibitem{fazel:04}M.~Fazel, H.~Hindi, and S.~Boyd, Proc. American Control Conference, Boston,
Massachusetts (2004).

\bibitem{mesbahi:97}M.~Mesbahi and G. P.~Papavassilopoulos, IEEE
Trans. on Automatic Control {\bf 42}, 23943 (1997).

\bibitem{Yalmip:04}
J.~Lofberg.
\newblock Yalmip: A toolbox for modeling and optimization in {MATLAB}.
\newblock In {\em Proceedings of the CACSD Conference}, Taipei, Taiwan, 2004.

\bibitem{Sdpt3}
K.~C. Toh, R.~H. Tutuncu, and M.~J. Todd.
\newblock Sdpt3: Matlab software for semidefinite-quadratic-linear programming.
\newblock 2004.
\newblock http://www.math.nus.edu.sg/$\sim$mattohkc/sdpt3.html.

\bibitem{Feynman:QC1} R.P. Feynman, Found. Phys. {\bf 16}, 507 (1986).
  
\end{thebibliography}

\appendix
\section{Proof of Equations (9), (17), (19)}

\label{sec:maxvr01}

The $\nc\times\nc\nra$ matrix $W=E(\Delta^\dag\otimes C)$ has a
maximum rank of $\nc$. Hence a singular value decomposition is of the
form $W=USV^\dag,\ S=[S_0\ 0], $ with $S_0$ an $\nc\times\nc$ diagonal
matrix containing the $\nc$ singular values. If $V$ is partitioned as
$V=[V_1\ V_2]$ with $V_1$ being $\nra\nc\times\nc$ then the objective
function in \eqref{maxvr} becomes, \begin{equation} \real\ \trace\ RW
= \real\ \trace\ S_0 X,\ X=V_1^\dag RU \end{equation} Since
$\norm{X}\leq 1$, then $\real\ \trace\ S_0 X \leq \trace\
S_0$. Equality occurs if and only if $X=\Ic$, or equivalently,
$R=V_1U^\dag$, which is precisely the result in \eqref{ropt ind}. This
also establishes that the optimal objective function is $\trace\ S_0$
which, by definition, is equal to $\trace\sqrt{WW^\dag}$, thus
\begin{equation}\label{sqrtww} \max_{R^\dag R=\Ic} \real\trace\ RW =
\trace\sqrt{WW^\dag} \end{equation} which establishes \eqref{maxvr}.

Condition \eqref{klgen} follows directly from \eqref{ind} by
multiplying both sides by their respective conjugate (with indices $c$
and $c^{\prime }$) which also eliminates $R$ because $R^\dag
R=\Ic$. This immediately establishes that \eqref{klgen} is a
\emph{necessary condition} for \eqref{ind}. To prove
\emph{sufficiency}, first expand \eqref{dind} to get,
\begin{equation}
  \label{osreq}
  \begin{array}{rcl} \dind
&=& \sum_c \trace\ (\Ie\otimes C_c^\dag)E^\dag E(\Ie\otimes C_c) \\
&& +\trace\ \Gamma_c\otimes\Is
-2\real\trace\ RE(\Delta_c^\dag\otimes C_c\udes^\dag) \\
\Gamma_c &=& \Delta_c^\dag\Delta_c
  \end{array}
\end{equation}
From \eqref{sqrtww}, we get,
\begin{equation}
\begin{array}{rcl} \min_{R^\dag R=\Ic}\dind
&=& \sum_c[ \trace\ (\Ie\otimes C_c^\dag)E^\dag E(\Ie\otimes C_c) \\
&& +\trace\ \Gamma_c\otimes\Is
] -2\trace\sqrt{WW^\dag} \\
  W &=& \sum_c\ E(\Delta_c^\dag\otimes C_c\udes^\dag)
\end{array}
\end{equation}
Using \eqref{klgen} we get, $\trace\sqrt{WW^\dag} = \sum_c\ E(\Ie
\otimes C_c C_c^\dag)E^\dag $. This, together with repeated uses of
\eqref{klgen} shows that $\min_{R^\dag R=\Ic}\dind=0$. Since $\dind$
is a norm, and is zero, then so is its argument, which by definition
establishes \eqref{ind} and thus shows sufficiency of \eqref{klgen}.

\section{Relation between fidelity $\favg$ and distance $\dind$}
\label{sec:delta}

The problem is,
\begin{equation}
  \begin{array}{ll} 
\mathrm{minimize} & 
\dind=\sum_c\ \normf{ R E(\Ie\otimes C_c)-\Delta_c\otimes \udes }^2
\\
\mathrm{subject~to} & \sum_c \normf{\Delta_c}^2=1
  \end{array}
\end{equation}
Form the Lagrangian,
\begin{equation}
L = \dind + \lam(\sum_c \trace\ \Delta_c^\dag\Delta_c-1)
\end{equation}
with $\lam$ the Lagrange multiplier.  Then, $\nabla_{\delta_{rec}}L=0$
when $(\ns+\lam)\delta_{rec}=\trace\ R_rE_eC_c\udes^\dag$.  To enforce
the constraint $\sum_c\normf{\Delta_c}^2=1$ requires that
$(\ns+\lam)^2=\sum_{r,e,c}|\trace\ R_rE_eC_c\udes^\dag|^2$.  Hence,
\beq
\bea{rcl}
\Delta_c
&=&
\bar{\Delta}_c / \sqrt{ \sum_c\normf{\bar{\Delta}_c}^2 }
\\
\bar{\delta}_{rec} &=& \trace(R_rE_eC_c\udes^\dag)/\ns 
\eea
\eeq
Observe that $\sum_c\normf{\bar{\Delta}_c}^2=f$. This together with
$\sum_rR_r^\dag R_r=\Ic,\ \sum_cC_c^\dag C_c=\Is$ gives the optimal
distance as given implicitly by \refeq{fdrel}. Note also that with no
constraint, $\lam=0$, the $\bar{\Delta}_c$ are the optimal
least-squares (unconstrained) solution.

\section{Unitary freedom in Equation (17)}

\label{sec:osr free}

In \eqref{maxvr}, $\Gamma=\Delta^\dag\Delta$ remains unchanged if
$\Delta$ is multiplied by a unitary. This unitary freedom is exactly
the unitary freedom
in describing the error map OSR. To see this, recall again from
\cite[Thm.8.2]{NielsenC:00} that two error maps with OSR elements 
$E=[E_1 \ldots E_\me]$ and $F=[F_1 \ldots F_\me]$ are equivalent if
and only if $E_i = \sum_j W_{ij} F_j$ where the $\me\times\me$
matrix $W$ is unitary. Equivalently from \eqref{osreq},
$E=F(W\otimes\Ic)$. Substituting this for $E $ into the left hand
side of \eqref{maxvr} gives, \begin{equation}
\real\ \trace\ RE(\Delta^\adj\otimes C \udes^\adj) = \real\ \trace\
RF((\Delta')^\adj\otimes C \udes^\adj) \end{equation} with
$\Delta^{\prime }= \Delta W^\dag$. Hence, $\Delta^{\prime \dag}\Delta^{\prime
}=W\Delta^\dag\Delta W^\dag=\Gamma$, which establishes the claim.

\section{Solving Equation (20)
via an SDP}

\label{sec:gamopt}

Problem \eqref{optgamma} is of the form,
\begin{equation}
  \label{gamopt p}
  \begin{array}{ll} 
\mathrm{maximize} & \trace\sqrt{F(\Gamma)} \\
\mathrm{subject~to} & \Gamma\geq 0,\ \trace\ \Gamma =1
  \end{array}
\end{equation}
where $F(\Gamma)$ is linear in $\Gamma$. Consider the relaxed problem,
\begin{equation}
  \label{gamopt prlx}
  \begin{array}{ll}
    \mathrm{maximize} & \trace\ Y \\
\mathrm{subject~to} & F(\Gamma) - Y^2 \geq 0,\ \Gamma\geq 0,\
\trace\ \Gamma =1
\end{array}
\end{equation}
This is an SDP in $\Gamma$ and $Y$ with Lagrangian,
\begin{equation}\label{lag prlx} \begin{array}{rcl}
L(\Gamma,Y,P,Z) &=& -\trace\ Y -\trace\ P(F(\Gamma)-Y^2) \\
&& - \trace\ Z\Gamma
+\lam(\trace\ \Gamma -1) \end{array}
\end{equation}
The dual function is, \begin{equation}\label{dual grlx}
  \begin{array}{rcl} g(P,\lam,Z) &=& \inf_{\Gamma,Y} L(\Gamma,Y,P,Z) \\
&=& \left\{ \begin{array}{ll} \inf_Y \trace(PY^2-Y)-\lam, & Z = \lam I - A(P) \\
-\infty & \mathrm{otherwise} \end{array} \right. \eea \end{equation}
with $A(P)=\frac{\partial}{\partial\Gamma}\trace\ PF(\Gamma)$, which
is not dependent on $\Gamma$ because $F(\Gamma)$ is linear in
$\Gamma$. Performing the
indicated $\inf_Y$ gives $Y=(1/2)P^{-1}$ and $g=-(\lam+(1/4)\trace\
P^{-1})$. The dual optimization associated with \eqref{gamopt prlx} is
to maximize $g $, or equivalently, minimize its negative, \ie,
\begin{equation}\label{gamopt drlx} \begin{array}{ll}
\mathrm{minimize} & \lam + \frac{1}{4}\trace\ P^{-1} \\
\mathrm{subject~to} & P>0,\ \lam I - A(P) \geq 0 \end{array}
\end{equation}
This is an SDP in the dual variables $P,\ \lam$. For this problem
\emph{strong duality} holds \cite{BoydV:04}. Consequently, at
optimality of \eqref{gamopt prlx} and \eqref{gamopt drlx} the
complementary slackness condition is
$P_\opt(F(\Gamma_\opt)-Y_\opt^2)=0$. Since $P_\opt>0$, we have $Y_
\opt = \sqrt{F(\Gamma_\opt)}$. This establishes that solving the SDP
\eqref{gamopt prlx} is equivalent to solving the original problem
\eqref{gamopt p}. 

\section{Solving for $C$ in Equation (24)}
\label{sec:copt}

The problem is,
\beq
\bea{ll}
\mbox{minimize} & \dind
=
\sum_c\ \normf{ R E(\Ie\otimes C)-\Delta\otimes \udes }^2
\\
\mbox{subject to} & C^\dag C=\Is
\eea
\eeq
Form the Lagrangian,
\beq
L = \dind + \trace\ P(C^\dag C-\Is)
\eeq
with $P$ the Lagrange multiplier.  Then, $\nabla_CL=0$ when
$C=\Cb(\Is+P)^{-1}$ with $\Cb$ as defined in \eqref{copt}.  To enforce
the constraint $C^\dag C=\Is$ requires that
$(\Is+P)^2=\Cb^\dag\Cb$. Hence, $C=\Cb(\Cb^\dag\Cb)^{-1/2}$.  The
actual computation of $C$ is done using the SVD \eqref{copt
svd1}-\eqref{copt svd2}.  Note that with no constraint, $P=0$, and
$\Cb$ is the optimal least-squares (unconstrained) solution.

\end{document}